\documentclass[authoryear]{elsarticle}
\usepackage{natbib}
\usepackage{enumerate}
\usepackage{amsmath,amsthm}
\usepackage{amssymb, latexsym}
\usepackage[mathscr]{euscript}
\usepackage{color}
\usepackage{array}
\usepackage{flafter}
\usepackage{layout}
\usepackage{graphicx}
\usepackage[colorlinks=true,linkcolor=black, citecolor=blue, urlcolor=blue, pdfauthor=author]{hyperref}
\usepackage{epstopdf}
\usepackage[ansinew]{inputenc}
\usepackage{xfrac}
\usepackage{algorithm}
\theoremstyle{plain}
  \newtheorem{teo}{Theorem}

\theoremstyle{definition}
  
  \newtheorem{ej}{Example}
\theoremstyle{remark}
  \newtheorem{remark}{Remark}

  \makeatletter
\def\ps@pprintTitle{%
   \let\@oddhead\@empty
   \let\@evenhead\@empty
   \def\@oddfoot{\reset@font\hfil\thepage\hfil}
   \let\@evenfoot\@oddfoot
}
\makeatother

\makeatletter
\providecommand{\doi}[1]{%
  \begingroup
    \let\bibinfo\@secondoftwo
    \urlstyle{rm}%
    \href{http://dx.doi.org/#1}{%
      doi:\discretionary{}{}{}%
      \nolinkurl{#1}%
    }%
  \endgroup
}
\makeatother

\title{Active information, missing data and prevalence estimation}

\author[su]{Ola H\"ossjer}\ead{ola@math.su.se}

\author[bst]{Daniel Andr\'es D\'{\i}az--Pach\'on\corref{cor1}}\ead{Ddiaz3@miami.edu}

\author[bst]{Chen Zhao\corref{cor2}}\ead{cxz511@miami.edu}

\author[bst]{J. Sunil Rao\corref{cor2}}\ead{JRao@miami.edu}

\cortext[cor1]{Corresponding author}
\address[su]{Department of Mathematics, Stockholm University, Roslagsv\"agen 101, Kr\"aftriket, hus 6 Room 318, Stockholm, Sweden}
\address[bst]{Division of Biostatistics - University of Miami, Don Soffer Clinical Research Center, 1120 NW 14th St, Miami FL, 33136}

\begin{document}

\begin{abstract}
The topic of this paper is prevalence estimation from the perspective of active information. Prevalence among tested individuals has an upward bias under the assumption that individuals' willingness to be tested for the disease increases with the strength of their symptoms. Active information due to testing bias quantifies the degree at which the willingness to be tested correlates with infection status. Interpreting incomplete testing as a missing data problem, the missingness mechanism impacts the degree at which the bias of the original prevalence estimate can be removed. The reduction in prevalence, when testing bias is adjusted for, translates into an active information due to bias correction, with opposite sign to active information due to testing bias. Prevalence and active information estimates are asymptotically normal, a behavior also illustrated through simulations.  
\end{abstract}


\maketitle

\section{Introduction}

According to the No Free Lunch Theorems, in a search problem, on average, no search does better than blind \citep{WolpertMacReady1997}. Therefore, when for a particular case one search does different than a uniform search (better or worse), it is because the programmer used her knowledge (good or bad) either of the target or the structure of the space, or both. Active information was introduced to measure the amount of information a programmer infuses in a search to reach the target with different probability than through a blind search \citep{DembskiMarks2009a, DembskiMarks2009b}. For a search space $\mathcal X$ and a target $A \subset \mathcal X$, active information is then naturally defined as $I^+ = \log (p/p_0)$, where $p$ is the probability of reaching $A$ under the algorithm devised by the programmer, and $p_0$ is the uniform probability of reaching $A$.

Another interpretation of active information will allow to see that a data set in $\mathcal X$, whose distribution is consistent with a probability $p$ of reaching $A$, will have a local mode in the region $A$ if $I^+> 0$ \citep{DiazEtAl2019, LiuEtAl2022}. Monta\~nez and collaborators have also used active information to analyze intention perception  \citep{HomEtAl2021}.  \citet{HossjerDiaz2022} have used active information to measure fine-tuning. And \citet{DiazMarks2020b} used it to compare non-neutral to neutral population genetics models. 

In this paper active information is used to unify estimation and bias correction of the prevalence $p_0$ of a disease when data is missing. This corresponds to a setting where $\mathcal X$ is a population of individuals whereas $A$ is the subpopulation of affected individuals. It is assumed that such a prevalence estimate is computed from a subsample of tested individuals and that the data analyst does not control the sampling scheme, but rather that individuals voluntarily choose to be tested. Since individuals with stronger symptoms are more likely to have the disease and get tested, knowledge of these symptoms represents information that leads to an estimated prevalence with an upward bias $p-p_0$. This bias is quantified in terms of a positive active information $I_T^+=\log(p/p_0)$ due to testing bias, since it quantifies the degree at which individuals' willingness to be tested correlates with their symptoms.  

Incomplete testing is regarded as a missing data problem \citep{LittleRubin2002}, and various missingness mechanisms will be considered here. In particular, when data is missing at random (\textsc{mar}), the bias of the prevalence estimate can be removed. This corresponds to a negative active information $I_C^+$ due to bias correction. Under ideal \textsc{mar} conditions, when the bias correction is successful, the total active information $I^+ = I^+_T + I^+_C$, after bias correction, is zero.

\section{Active information due to testing bias}\label{Sec:Prev}

Let $\mathcal X$ be a population of $N=|\mathcal X|$ individuals of which those in $A\subset\mathcal X$ have a certain disease, whereas the other subjects in $A^c=\mathcal X\setminus A$ have not. Let $P_0$ refer to the uniform probability measure on $\mathcal X$, which assigns a probability of $1/N$ to each individual. The objective is to estimate the population prevalence
\begin{equation}\label{p0}
	p_0 = P_0(A) = \frac{|A|}{N}
\end{equation}
of the disease from a subgroup of individuals that are tested. To this effect, first divide 
\begin{equation}\label{cXDecomp}
	\mathcal X = \cup_{s=0}^{S-1}\cup_{i=0}^1 \mathcal X_{si},
\end{equation} 
into a number of subpopulations of unknown sizes $|\mathcal X_{si}|=N\rho_{si}$, where $\mathcal X_{si}$ consists of those individuals with symptoms $s\in\{0,\ldots,S-1\}$ and infection status $i\in\{0,1\}$. The first variable $s$ is measured on an ordinal scale with increasingly stronger symptoms, so that $s=0$ represents no symptoms whereas $s=S-1$ codes for the strongest possible symptoms. Infection status, on the other hand, is a binary variable such that $i=0$ and $i=1$ correspond to a non-infected and infected individual, respectively. For each $x\in\mathcal X$ we let
\begin{equation}\label{Ix}
	I(x)\in \{(0,0),\ldots,(S-1,0),(0,1),\ldots,(S-1,1)\}
\end{equation}
signify the subpopulation $\mathcal X_{si}$ to which $x$ belongs. 

Let also $T_x$ be a variable that equals 1 or 0 depending on whether $x$ is tested for the disease or not. The collection $\{T_x;\, x\in\mathcal X\}$ is assumed to be formed by independent Bernoulli variables, with 
$P(T_x=1)=\pi_{I(x)}$. This corres\-ponds to an assumption whereby individuals in different groups are tested with different sampling probabilities $\pi_{si}$. Consequently, the weighted probability measure 
\begin{equation}\label{Px}
	P(x) = \frac{\pi_{I(x)}}{\sum_{y\in\mathcal X} \pi_{I(y)}}, \quad x\in\mathcal X
\end{equation}
represents a prediction of the tested population, before testing has occurred. In particular, the testing prevalence
\begin{equation}\label{p}
	p = P(A) = \sum_{x\in A} P(x)
\end{equation}
is the expected prevalence in the tested subpopulation. The active information due to testing bias is defined as 
\begin{equation}\label{I+T}
	I^+_T = \log \frac{p}{p_0} = \log \frac{P(A)}{P_0(A)}.
\end{equation}
To estimate $p$ and $I^+_T$, the subpopulation
\begin{equation}\label{cXT}
	\mathcal X_T = \{x\in\mathcal X; \, T_x=1\}
\end{equation}
of $N_T=|\mathcal X_T|$ tested individuals is introduced. Since $N_T$ is known, this gives rise to an estimator
\begin{equation}\label{hp}
	\hat{p} = \hat p(A) = \frac{|A\cap \mathcal X_T|}{N_T}
\end{equation}
of $p$. The expected fraction of sampled individuals is also introduced as
\begin{equation}\label{pi}
	\pi = \sum_{i=0}^1\sum_{s=0}^{S-1} \rho_{si}\pi_{si},
\end{equation}
which is estimated by 
\begin{equation}\label{hpi}
	\hat \pi = \frac{N_T}{N}.
\end{equation}

\section{Active information after bias correction}\label{Sec:ActInfo}

The relation between $p$ and $p_0$ depends crucially on the sampling probabilities $\pi_{si}$. This can be seen by noting that the population and testing prevalences are different functions
\begin{equation}\label{p0p}
		p_0 = \sum_{s=0}^{S-1} \rho_{s1},\quad p = \sum_{s=0}^{S-1} \rho_{s1}\pi_{s1}/\sum_{s,i}  \rho_{si}\pi_{si}
\end{equation}
of $\rho_{01},\ldots,\rho_{S-1,1}$. Regarding non-tested individuals as missing data, concepts from the missing data literature \citep{LittleRubin2002} are helpful to explain the way in which data is missing. Random sampling, or data missing completely at random (\textsc{mcar}), occurs when 
\begin{equation}\label{MCAR}
	\pi_{si} = \pi.
\end{equation}
From \eqref{p0p}, $p=p_0$ and $I_T^+=0$ whenever \eqref{MCAR} holds. Condition \eqref{MCAR} is usually very unrealistic, since people with stronger symptoms (larger $s$) are more likely to be tested (have larger $\pi_{s0}$ and $\pi_{s1}$) than those with weaker symptoms. A weaker assumption of data missing at random (\textsc{mar}) occurs when the sampling probabilities only depend on variables that are known. In an example of a \textsc{mar}~sampling scheme $\rho_s = \rho_{s0}+\rho_{s1}$ is known and 
\begin{align}\label{MAR}
		\pi_{si} = \pi_s
\end{align}
for $s=0,\ldots,S-1$. The most challenging missingness mechanism (neither \textsc{mcar}~or \textsc{mar}) is referred to as data missing not at random (\textsc{mnar}). 

Also from \eqref{p0p}, typically $p\ne p_0$ and $I^+_T\ne 0$ when data is \textsc{mar}~or \textsc{mnar}. To construct a bias-corrected estimator $\hat{p}_0$ of $p_0$, the biased prevalence estimator can be rewritten as
\begin{equation}\label{hp2}
	\hat p = \frac{\sum_{s=0}^{S-1} \rho_{s1}\tilde \pi_{s1}}{\sum_{s,i} \rho_{si}\tilde \pi_{si}} = \sum_{s=0}^{S-1} \rho_{Ts1},
\end{equation}
where the sampling fractions
\begin{equation}\label{tpisi}
	\tilde \pi_{si} = \frac{|\mathcal X_T\cap \mathcal X_{si}|}{|\mathcal X_{si}|} = \frac{|\mathcal X_{Tsi}|}{|\mathcal X_{si}|} = \frac{N_{Tsi}}{N_{si}} = \frac{N_{Tsi}}{N\rho_{si}} 
\end{equation}
for different subpopulations approximate $\pi_{si}$, whereas 
\begin{equation}\label{rhoTsi}
	\rho_{Tsi} = \frac{\rho_{si}\tilde \pi_{si}}{\sum_{r,k} \rho_{rk}\tilde \pi_{rk}} = \frac{N_{Tsi}}{N_T} 
\end{equation}
are the known fractions at which the subpopulations appear in the sample. A comparison between \eqref{p0p} and \eqref{rhoTsi} suggests an estimate 
\begin{equation}\label{hp0}
	\hat p_0 = \frac{\sum_{s=0}^{S-1} \rho_{Ts1}\hat \pi_{s1}^{-1}}{\sum_{s,i} \rho_{Tsi}\hat \pi_{si}^{-1}}
		= \frac{\sum_{s=0}^{S-1} N_{Ts1}\hat \pi_{s1}^{-1}}{\sum_{s,i} N_{Tsi}\hat \pi_{si}^{-1}}
\end{equation}
of the population prevalence $p_0$, where $\hat \pi_{si}$ is an estimate of $\tilde \pi_{si}$ (and thereby also an estimate of $\pi_{si}$). Plugging \eqref{hp} and \eqref{hp0} into \eqref{I+T}, the estimator 
\begin{equation}\label{hI+} 
	\hat I^+_T = \log \frac{\hat p}{\hat p_0} 
\end{equation}
of the active information $I^+_T$ due to testing bias is obtained.  Furthermore, 
\begin{equation}\label{I+}
	I^+ = \log \frac{E(\hat p_0)}{p_0} = \log \frac{p}{p_0} + \log \frac{E(\hat p_0)}{p} = I_T^+ + I_C^+
\end{equation}
will be referred to as the active information of the bias-adjusted prevalence estimate \eqref{hp0}, which is a sum of two terms: the active information \eqref{I+T} due to testing bias and the active information $I_C^+$ due to bias correction. If the bias correction is completely successful ($I^+=0$), then $I_C^+ = -I_T^+$. This suggests an estimate
\begin{equation}\label{hI+C}
	\hat I_C^+ = - \hat I_T^+ = -\log \frac{\hat p}{\hat p_0}
\end{equation}
of $I_C^+$.

\begin{ej}[\textsc{mcar}] 
Whenever \eqref{MCAR} holds, $I_T^+=I_C^+=0$ follows from \eqref{I+T} and \eqref{p0p}. In this context, to assume that the estimated sampling fractions $\hat \pi_{si}^{\textsc{mcar}}=\hat \pi^{\textsc{mcar}}$ are the same for all subpopulations $\mathcal X_{si}$ is natural. Since $\hat \pi^{\textsc{mcar}}$ cancels out in the prevalence estimator \eqref{hp0}, it simplifies to 
\begin{equation}\label{hp0MCAR}
	\hat p_0^{\textsc{mcar}} = \frac{\sum_{s=0}^{S-1} \rho_{Ts1}}{\sum_{s,i} \rho_{Tsi}} = \sum_{s=0}^{S-1} \rho_{Ts1}
	= \frac{1}{N_T}\sum_{s=0}^{S-1} N_{Ts1} = \frac{N_{T\cdot 1}}{N_T} = \hat p,
\end{equation}
and consequently $\hat I_T^+ = \hat I_C^+ = 0$ under \textsc{mcar}~sampling. From Fisher's exact test, $N_{T\cdot 1}$ has a hypergeometric distribution
\begin{equation}\label{NTcdot1Distr}
	N_{T\cdot 1} \mid N_T \sim \text{Hyp}(N,N_T,p_0)
\end{equation}
conditionally on $N_T$. Taking expectations in both sides of \eqref{hp0MCAR}, by  \eqref{rhoTsi} and \eqref{hp2}, $E\left(\hat p_0^{\textsc{mcar}}\right) = p_0$ and $I^+ = 0$.  
\end{ej}

\begin{ej}[\textsc{mar}]\label{Exa:MAR} 
The \textsc{mar}~sampling scheme \eqref{MAR} can be viewed as an instance of stratified sampling \citep{GrovesEtAl2009}, where the relative sizes $\rho_s$ of the strata (symptom classes) are known. Although the sampling fractions $\tilde \pi_{si}$ in \eqref{tpisi} are unknown when \eqref{MAR} holds, they may be estimated consistently by means of  
\begin{equation}\label{hpisMAR}
	\hat \pi_{si}^{\textsc{mar}} = \hat \pi_s^{\textsc{mar}} = \frac{N_{Ts0}+N_{Ts1}}{N\rho_{s0}+N\rho_{s1}} = \frac{N_{Ts}}{N\rho_s},
\end{equation}
where in the last step $N_{Ts}=N_{Ts0}+N_{Ts1}$ was introduced. Plugging \eqref{hpisMAR} into \eqref{hp0}, the estimator 
\begin{equation}\label{hp0MAR}
	\hat p_0^{\textsc{mar}} =  \frac{\sum_{s=0}^{S-1} \rho_{Ts1}(N\rho_s/N_{Ts})}{\sum_{s,i} \rho_{Tsi}(N\rho_s/N_{Ts})} 
	= \frac{\sum_{s=0}^{S-1} \rho_s (N_{Ts1}/N_{Ts})}{\sum_{s=0}^{S-1} \rho_s} 
	= \sum_{s=0}^{S-1} \rho_s \hat p_{0s}
\end{equation}
of $p_0$ is obtained. It is a weighted average of estimates
\begin{equation}\label{hp0s}
	\hat p_{0s} = \frac{N_{Ts1}}{N_{Ts}}
\end{equation}
of the prevalences
\begin{equation}\label{p0s}
	p_{0s} = \frac{\rho_{s1}}{\rho_s}
\end{equation}
in symptom classes $\mathcal X_s = \mathcal X_{s0}\cup \mathcal X_{s1}$, using data from cohorts $\mathcal X_{Ts} = \mathcal X_{Ts0}\cup \mathcal X_{Ts1}$. Since $\pi_{s0}=\pi_{s1}=\pi_s$, from Fisher's exact test,
\begin{equation}\label{NTs1Distr}
	N_{Ts1} \mid N_{Ts} \sim \text{Hyp}(N_s,N_{Ts},p_{0s})
\end{equation}
for $s=0,\ldots,S-1$. In view of \eqref{p0p}, this implies 
\begin{equation}\label{Ehp0MAR}
	E\left(\hat p_0^{\textsc{mar}}\right) = \sum_{s=0}^{S-1} \rho_s E(\hat p_{0s}) =  \sum_{s=0}^{S-1} \rho_{s}p_{0s} = p_0.
\end{equation}
Consequently, $I^+=0$ under \textsc{mar}~sampling, although in general $I_T^+ = - I_C^+$ differs from zero.   
\end{ej}

\begin{ej}[A model for COVID-19 testing]\label{Exa:COVID} 
In the context of COVID-19 testing, \citet{DiazRao2021} and \cite{ZhouDiazRao2022} considered a model of convenience sampling with $S=2$ symptom classes such that
\begin{align}\label{piCOVID}
		\pi_{00} = \pi_{01} &= \pi_0,\quad   \pi_{10} = \pi_{11} = \pi_1,\quad \pi_1 > \pi_0.
\end{align}
Since $\rho_0$ and $\rho_1$ are unknown, this is not a \textsc{mar}~model in the sense of the condition \eqref{MAR}.  In fact, the third assumption of \eqref{piCOVID} says that in convenience sampling symptomatic individuals are more likely to get tested than asymptomatic ones, which implies that with high probability $\rho_1 < N_{T1}/N_T$. On the other hand, the presence of $N_{T1}$ symptomatic individuals in the sample implies that $N_{T1}/N \le \rho_1$. 

From a Bayesian approach and the maximum entropy principle \citep{Jaynes1968, DiazMarks2020a},  $\rho_1$ is then assumed to be uniformly distributed inside the interval $(N_{T1}/N, N_{T1}/N_T)$. Therefore, $\hat \rho_1 = E(\rho_1)$ is taken as the estimator of the proportion of symptomatic individuals in the population, and $\hat \rho_0 = 1 - \hat \rho_1 = E(\rho_0)$ estimates the proportion of the asymptomatic group. From this viewpoint, a modification of \eqref{hpisMAR} produces
\begin{equation}\label{hpiCOVID}
	\hat \pi_{si}^{\text{MaxEnt}} = \hat \pi_s^{\text{MaxEnt}} =  \frac{N_{Ts}}{NE(\rho_s)},
\end{equation}
and plugging \eqref{hpiCOVID} into \eqref{hp0}, the estimator of prevalence is 
\begin{align}\label{hp0COVID}
	\hat p_0 &= \frac{N_{T01}(\hat \pi_0^{\text{MaxEnt}})^{-1} + N_{T11}(\hat \pi_1^{\text{MaxEnt}})^{-1}}{N_{T0}(\hat \pi_0^{\text{MaxEnt}})^{-1} + N_{T1}(\hat \pi_1^{\text{MaxEnt}})^{-1}} \nonumber \\
			&= \frac{N_{T01}}{N_{T0}}(1-\hat \rho_1) + \frac{N_{T11}}{N_{T1}}\hat \rho_1,
\end{align}
where \eqref{hp0COVID} is obtained using the first two assumptions of \eqref{piCOVID}, which imply that inside each group of symptoms the sampling of infected and non-infected is random.
\end{ej}

\section{Asymptotics}\label{Sec:AsPrevEst}

This section is focused on the asymptotic properties of the estimates $\hat p$ and $\hat p_0$ of the test-biased and population-based prevalences $p$ and $p_0$, as the population size gets large ($N\to\infty$). The second part of the \textsc{mar}~condition is assumed:
\begin{equation}\label{psipi}
	\pi_{si}=\pi_s, \quad s=0,\ldots,S-1,
\end{equation}
In conjunction with \eqref{p0p} and \eqref{p0s}, this makes possible to rewrite the expected prevalence among the tested individuals as 
\begin{equation}\label{p2}
	p = \sum_{s=0}^{S-1} \tilde \rho_s p_{0s},
\end{equation}
where 
\begin{equation}\label{trhosi}
	\tilde \rho_{s} = \frac{\rho_s \pi_s}{\sum_{r=0}^{S-1} \rho_r \pi_r}.
\end{equation}
is the expected proportion of tested individuals with symptoms $s$. The estimator of $p$ can equivalently be expressed as 
\begin{equation}
	\hat p = \sum_{s=0}^{S-1} \rho_{Ts} \hat p_{0s},
\end{equation}
where 
\begin{equation}\label{rhoTs}
	\rho_{Ts} = \frac{N_{Ts}}{N_T} = \rho_{Ts0} + \rho_{Ts1}
\end{equation}
is an estimate of $\tilde \rho_s$. 

In view of \eqref{psipi}, the requirement is made that $\hat \pi_{si}=\hat \pi_s$, and 
\begin{equation}\label{hrhos}
	\hat \rho_s = \frac{N_{Ts}\hat \pi_s^{-1}}{\sum_{r=0}^{S-1} N_{Tr}\hat \pi_r^{-1}},
\end{equation}
is introduced, so that the bias-corrected prevalence estimator in \eqref{hp0} simplifies to 
\begin{equation}\label{hpi02}
	\hat p_0 = \sum_{s=0}^{S-1} \hat \rho_s  \hat p_{0s}.   
\end{equation}
The quality of $\hat p_0$ as an estimator of $p_0$ in \eqref{Ehp0MAR}, depends on how well $\hat \rho_s$ estimates $\rho_s$. Therefore,
\begin{equation}\label{bap0N}
	\bar p_{0N} = E\left(\hat p_0|\{\hat \rho_s\}_{s=0}^{S-1}\right) = \sum_{s=0}^{S-1} \hat \rho_s p_{0s}
\end{equation}
is introduced.

The following theorem provides the asymptotic properties of $\hat p_0$, $\hat p$, and $\hat I_T^+$: 

\begin{teo}\label{Thrm:As}
Suppose $N\to\infty$ in such a way that $\rho_s = N_s/N$ are kept fixed, that \eqref{psipi} holds for fixed $\pi_1,\ldots,\pi_s$ and
\begin{equation}\label{hrhosConv}
	\hat \rho_s \longrightarrow_p \bar{\rho}_s
\end{equation}
as $N\to\infty$, for $s=0,\ldots,S-1$, where $\longrightarrow_p$ refers to convergence in probability. Then $\hat p$, $\hat p_0$, $\hat I_T^+$ are asymptotically normally distributed as $N\to\infty$, in the sense that
\begin{equation}\label{hpConv}
	N^{1/2} (\hat p-p) \longrightarrow_\mathcal L N(0,V_1+V_2),
\end{equation}
\begin{equation}\label{hp0Conv}
	N^{1/2} (\hat p_0-\bar p_{0N}) \longrightarrow_\mathcal L N(0,V_3),
\end{equation}
and 
\begin{equation}
	N^{1/2} \left[\hat I_T^+ - \left(I_T^+ - \log \frac{\bar p_{0N}}{p_0}\right)\right] \longrightarrow_\mathcal L N\left(0,\frac{V_1+V_2}{p^2} + \frac{V_3}{\bar p_0^2} - \frac{2V_4}{p\bar p_0}\right)
\label{hI+TConv}
\end{equation}
where $\longrightarrow_\mathcal L$ refers to weak convergence as $N\to\infty$,
\begin{equation}\label{bap0}
	\bar p_0 = \sum_{s=0}^{S-1} \bar{\rho}_s p_{0s} 
\end{equation}
is the asymptotic limit of $\bar p_{0N}$ as $N\to\infty$ (i.e.\ $\bar p_{0N}\longrightarrow_p \bar p_0$), whereas $V_1$, $V_2$, $V_3$, and $V_4$ are defined in the proof of Theorem \ref{Thrm:As}. 
\end{teo}

\begin{remark}[Standard errors in confidence intervals]\label{RemCI} 
The asymptotic variances $\sigma_p^2=(V_1+V_2)/N$, $\sigma_{p_0}^2=V_3/N$, and $\sigma_{I_T^+}^2$ in formulas \eqref{hpConv}-\eqref{hI+TConv} are functions of $p_{0s}$, $\bar p_0$, $p$, $\rho_s$, $\bar{\rho}_s$, and $\pi_s$.  If estimates $\hat p_{0s}$, $\hat p_0$, $\hat p$, $\hat \rho_s$, $\hat \rho_s$, and $\hat \pi_s$ of these quantities are plugged into the asymptotic variances in \eqref{hpConv}-\eqref{hI+TConv}, it is possible to obtain standard errors $\hat \sigma_p$, $\hat \sigma_{p_0}$, and $\hat \sigma_{I_T^+}$ of $\hat p$, $\hat p_0$, and $\hat I_T^+$, respectively. The corresponding confidence interval of $I_T^+$, with asymptotic coverage probability $1-\alpha$, is
$$
\text{CI}_{I_T^+} = \left(\hat I_T^+ - \lambda_{\alpha/2}\hat \sigma_{I_T^+},\hat I_T^+ + \lambda_{\alpha/2}\hat \sigma_{I_T^+} \right),
$$
where $\lambda_{\alpha/2}$ is the $(1-\alpha/2)$-quantile of a standard normal distribution. The delta method is first used to determine confidence intervals for logit transformed versions $\text{logit}(p)=\log [p/(1-p)]$ and $\text{logit}(p_0)$ of the prevalence parameters \citep{Agresti2013, LehmannCasella1998}. A logistic back-transformation $\text{logit}^{-1}(z) = \exp(z)/(1+\exp(z))$ yields confidence intervals   
$$
\text{CI}_p = \left( \text{logit}^{-1}\left(\text{logit}(\hat p) - \frac{\lambda_{\alpha/2}\hat \sigma_p}{\hat p(1-\hat p)}\right),   
\text{logit}^{-1}\left(\text{logit}(\hat p) + \frac{\lambda_{\alpha/2}\hat \sigma_p}{\hat p(1-\hat p)}\right) \right)
$$
and
$$
\text{CI}_{p_0} = \left( \text{logit}^{-1}\left(\text{logit}(\hat p_0) - \frac{\lambda_{\alpha/2}\hat \sigma_{p_0}}{\hat p_0(1-\hat p_0)}\right),   
\text{logit}^{-1}\left(\text{logit}(\hat p_0) + \frac{\lambda_{\alpha/2}\hat \sigma_{p_0}}{\hat p_0(1-\hat p_0)}\right) \right)
$$
of $p$ and $p_0$ respectively, with approximate coverage probability $1-\alpha$. 
\end{remark}

\begin{remark}[\textsc{mar}]
Since $\rho_s$ is known, $\hat \rho_s=\bar{\rho}_s=\rho_s$ and $\bar p_{0N}=\bar p_0=p_0$ for the \textsc{mar}~sampling scheme. Then $\hat I_T^+$ is an asymptotically unbiased estimator of $I_T^+$, and \eqref{hI+TConv} simplifies to 
$$
N^{1/2}(\hat I_T^+ - I_T) \longrightarrow_\mathcal L N\left(0,\frac{V_1+V_2}{p^2} + \frac{V_3}{p_0^2} - \frac{2V_4}{p p_0}\right) 
$$
as $N\to\infty$. 
\end{remark}

\begin{remark}[A conditional version of active information] 
Suppose that the interest is in active information due to sampling bias {\it conditionally} on the number $N_{T0},\ldots,N_{T,S-1}$ of individuals with different symptoms that are tested. The corresponding prevalence and active information are 
\begin{equation}
\bar p = E\left(\hat p|\{N_{Ts}\}_{s=0}^{S-1}\right) = \sum_{s=0}^{S-1} \rho_{Ts} p_{0s}.
\label{bap}
\end{equation}
and 
\begin{equation}
\bar I_T^+ = \log \frac{\bar p}{p_0}
\label{baIT+}
\end{equation}
respectively. Using the same type of argument as in the proof of Theorem \ref{Thrm:As}, then 
\begin{equation}
N^{1/2}(\hat p - \bar p) \longrightarrow_\mathcal L N(0,V_1)
\label{hpbapConv}
\end{equation}
and 
\begin{equation}
N^{1/2}\left(\hat I_T^+ - \left(\bar I_T^+ - \log \frac{\bar p_{0N}}{p_0} \right)\right) \longrightarrow_\mathcal L N\left(0,\frac{V_1}{p^2} + \frac{V_3}{\bar p_0^2} - \frac{2V_4}{p\bar p_0}\right)
\label{hI+TConv2}
\end{equation}
as $N\to\infty$. The $V_2$ term is missing in \eqref{hpbapConv} and \eqref{hI+TConv2}, compared to \eqref{hpConv} and \eqref{hI+TConv}. This term corresponds to the fact that the actual proportions $\rho_{Ts}$ of tested individuals with different symptoms deviate slightly from the corresponding expected proportions $\tilde \rho_s$. Because of the missing variance terms of \eqref{hpbapConv} and \eqref{hI+TConv2}, the standard errors of $\hat p$ and $\hat I_T^+$ are smaller when a conditional rather than an unconditional approach is used, and the confidence intervals of $\bar p$ and $\bar I_T^+$ are shorter compared to those of $p$ and $I_T^+$.   
\end{remark}

\begin{ej}[Maximum entropy approach]
Consider a \textsc{mnar}~sampling scheme where the sizes $\rho_s$ of symptom classes are unknown, although lower and upper bounds $0\le a_{sN}\le \rho_s \le b_{sN}\le 1$ are known. The maximum entropy approach of Example \ref{Exa:COVID} is generalized assuming that the vector $\rho=(\rho_0,\ldots,\rho_{S-1})$ is a random variable supported on the set 
\begin{equation}\label{calrho}
	{\cal R} = \{\rho=(\rho_0,\ldots,\rho_{S-1}); \, a_{sN}\le \rho_s \le b_{sN}; \, \sum_{s=0}^{S-1} \rho_s = 1\},
\end{equation}
a subset of the $S$-dimensional simplex of dimension $0\le d \le S$. By the maximum entropy principle, $\rho$ has a uniform density $f_\rho$ on ${\cal R}$, which degenerates to a point mass at $\mathcal R$ when $d=0$. This gives rise to estimates 
\begin{equation}\label{hpisMaxEnt}
	\hat \pi_s = \frac{N_{Ts}}{NE(\rho_s)}
\end{equation}
of the sampling probabilities $\pi_s$. Inserting \eqref{hpisMaxEnt} into \eqref{hrhos}, 
\begin{equation}\label{hrhosMaxEnt}
	\hat \rho_s = \frac{E(\rho_s)}{\sum_{r=0}^{S-1} E(\rho_s)} = E(\rho_s) = \int_{\cal R} r_s f_\rho(r)dr,
\end{equation}
with $r=(r_0,\ldots,r_{S-1})$. In particular, the \textsc{mar}~sampling scheme of Example \ref{Exa:MAR} corresponds to the special case $a_{sN}=b_{sN}=E(\rho_s)=\rho_s=\hat \rho_s$ and $d=0$. 

For the COVID-19 model of Example \ref{Exa:COVID}, using \eqref{hpiCOVID} to rewrite \eqref{hrhos}, 
\begin{align}\label{hrhoCOVID}
		\hat \rho_0 = 1 - N_{T1}/(\hat \pi_1 N),\quad \hat \rho_1 &= N_{T1}/(\hat \pi_1 N).
\end{align}
With an apriori assumption $\hat \pi_0 \le \hat \pi_1$, equation \eqref{hrhoCOVID} implies that $\mathcal R$ has dimension $d=S=2$, since the conditions
\begin{align*}
		a_{0N} &= 1 - N_{T1}/N_T,\quad     b_{0N} = 1 - N_{T1}/N,\\
		a_{1N} &= N_{T1}/N,\quad     \text{\ \ \ \ \ \ \ \ \ } b_{1N} = N_{T1}/N_T
\end{align*}
imply
\begin{align}\label{abCOVID}
		\hat \rho_0  = 1 - \frac{N_{T1}}{2N_T}\left(\frac{N_T}{N} +1\right),\quad   \hat \rho_1 = \frac{N_{T1}}{2N_T}\left(\frac{N_T}{N} +1\right).
\end{align}
Insertion of \eqref{hp0s} and \eqref{abCOVID} into \eqref{hpi02} finally leads to \eqref{hp0COVID}.   
\end{ej}
 
\section{Numerical Illustrations}\label{Sec:Exa}

This section illustrates with simulations the methodology under the framework of Examples 1 and 2. In these simulations, $N$ denotes known population size which is increased from 1000 to 1000000.  The true population prevalence is set at $p_0 = 0.20$. Only two levels of symptoms will be considered, $s\in \{0,1\}$. The proportion of people with symptoms $\rho_1$ in the population is 0.20 and without symptoms, $\rho_0$ is 0.80.  The proportion of positive cases with symptoms $\rho_{11} = 0.15$, and the proportion of positive cases without symptoms $\rho_{01}= 0.05$. Notice that $\rho_{11}+\rho_{01} = p_0$. The testing group within each symptom class is also assumed to be independent of the disease condition, in accordance with (\ref{MAR}).  

Let $\pi_1$ be the probability of testing the symptomatic group, and $\pi_0$ be the probability of testing the asymptomatic group. In the case of \textsc{mcar}, $\pi_1 = \pi_0=\pi$ is set to $0.6$. Thus, the overall prevalence rate can be estimated by the positive rate (\ref{hp0MCAR}) in the testing sample.  

For the \textsc{mar}~scenario, the probability of testing in symptomatic group $\pi_1$ is set to $0.10$, while the probability of testing in the asymptomatic group is $\pi_0 = 0.90$. Thus, the estimated population prevalence is a weighted average of the positive test rate by proportion of testing.  

Finally an \textsc{mnar}~situation is also considered. Unlike \textsc{mar}, the simulations were repeated without assuming $\pi_{si} = \pi_{s}$. Here, $\pi_{00}$ = 0.20, $\pi_{01}$ = 0.30, $\pi_{10}$ = 0.70, $\pi_{11}$  = 0.80. Thus, using the weighted positive test rate as \textsc{mar}, biased results, for which the bias will not vanish asymptotically, are expected. Each experiment is repeated 500 times.

Table~\ref{table:mcarAI} shows the active information of \textsc{mcar}.  Here the probabilities were averaged over the 500 realizations before calculating the active information values.  
The estimated active information of the correction, $\hat I^+_C$, is 0 because $\hat{p}_0 = \hat{p}$ in \textsc{mcar}. Thus, the active information of the bias-adjusted prevalence estimate for \textsc{mcar}, $\hat I^+$, is obtained from $\hat{I}_T^+$. 

\begin{table}[h!]
\centering
 \caption{Empirical active information under \textsc{mcar}}
 \label{table:mcarAI}
 \begin{tabular}{||c c c c c||} 
 \hline
 Population & 1000 & 10000 & 100000 & 1000000 \\ [0.5ex] 
 \hline\hline
 $\hat I_T^+$ & 0.0022 & 0.00002 & 0.00003 & 0.00003 \\  [1ex] 
 $\hat I_C^+$ & 0 & 0 & 0 & 0 \\  [1ex] 
 $\hat I^+$ & 0.0022 & 0.00002 & 0.00003 & 0.00003 \\  [1ex] 
 \hline
 \end{tabular}
\end{table}

Next, active information values under the \textsc{mar} simulation were obtained, as shown in Table~\ref{table:marAI}. The active information of the bias-adjusted prevalence estimate in \textsc{mar}\ is seen to increase as population increases, removing asymptotically the effect of a small overcorrection. 

\begin{table}[h!]
\centering
 \caption{Empirical active information under \textsc{mar}}
 \label{table:marAI}
 \begin{tabular}{||c c c c c||} 
 \hline
 Population & 1000 & 10000 & 100000 & 1000000 \\ [0.5ex] 
 \hline\hline
 $\hat I_T^+$ & 0.9873 & 0.9901 & 0.9903 & 0.9905 \\  [1ex] 
 $\hat I_C^+$ & -0.9917 & -0.9914 & -0.9892 & -0.9907 \\  [1ex] 
 $\hat I^+$ & -0.0044 & -0.0013 & 0.0011 & -0.0002 \\  [1ex] 
 \hline
 \end{tabular}
\end{table}

For  \textsc{mnar}, $\hat I^+_T = \log \frac{E(\hat{p})}{p_0}$, and $\hat I^+_C = \log \frac{E(\hat{p}_0)}{E(\hat{p})}$. Then $\hat I^+ = \frac{E(\hat{p}_0)}{p_0}$.  The active information  of the bias-adjusted prevalence estimate for this simulation is shown in Table~\ref{table:mnarAI}, showing that the strategy partially corrects the sampling bias.

\begin{table}[h!]
\centering
 \caption{Empirical active information under \textsc{mnar}}
 \label{table:mnarAI}
 \begin{tabular}{||c c c c c||} 
 \hline
 Population & 1000 & 10000 & 100000 & 1000000 \\ [0.5ex] 
 \hline\hline
 $\hat I_T^+$ & 0.994 & 0.990 & 0.990 & 0.990 \\  [1ex] 
 $\hat I_C^+$ &-0.398 & -0.396 & -0.396 & -0.396 \\  [1ex] 
 $\hat I^+$ & 0.596 & 0.594 & 0.594 & 0.594 \\  [1ex] 
 \hline
 \end{tabular}
\end{table}

Empirical root mean squared errors (RMSE) for the bias-corrected population prevalence estimates under each scenario are reported in Table~\ref{table:1}, together with their standard deviations in parentheses.  Clearly empirical RMSEs drop to zero with increasing $N$ under \textsc{mar}~and \textsc{mcar}~but not under \textsc{mnar}~where even for very large population sizes, the estimation of population prevalence cannot be improved.

\begin{table}[ht!]
\centering
\caption{Empirical RMSE for three sampling model simulations}\label{table:1}
\begin{tabular}{cccc}
     & \multicolumn{3}{c}{Empirical RMSE (SD)} \\
    \cline{2-4}
    $N$ & \textsc{mcar} & \textsc{mar} & \textsc{mnar}\\ 
    \hline
    1000  & 0.0058 (0.0042) & 0.0218 (0.0129) & 0.164 (0.006) \\ 
    10000 & 0.0020 (0.0015)& 0.0072 (0.0045) & 0.162 (0.002) \\ 
    100000  &0.0006 (0.0005)& 0.0023 (0.0014)& 0.162 (0.001) \\ 
    1000000 &0.0002 (0.0001)& 0.0007 (0.0004) & 0.162 (0.001) \\ 
\end{tabular}
\end{table}

\subsection{Asymptotics}

Section \ref{Sec:AsPrevEst} develops the asymptotic limiting distribution of the bias-corrected population prevalence estimator $\hat{p}_0$ in (\ref{hpi02}). Two scenarios are explored here:  i) small $p_0$, where the proportion of symptomatic individuals in the population $\rho_1$ is set to $0.1$, and the population prevalence $p_0$ is set to $0.05$; and ii) large $p_0$, where $\rho_1 = 0.2$ and $p_0 = 0.15$.  

\begin{figure}
  \includegraphics[width=\linewidth]{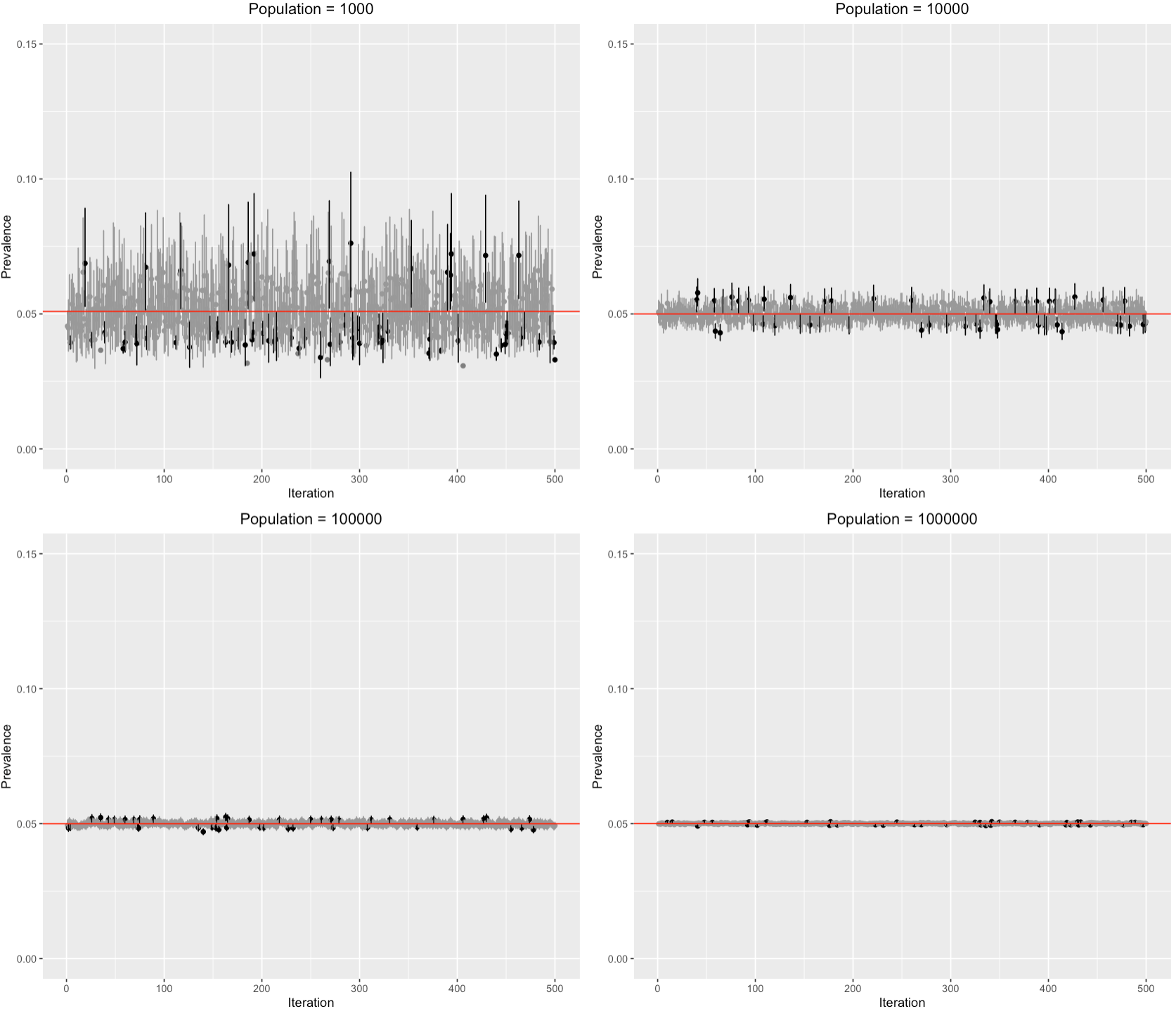}
  \caption{Confidence interval plots for 500 simulations and increasing population sizes under the \textsc{mar}~scenario 1, with $\rho_1=0.1$ and $p_0=0.05$.}
  \label{fig:smallCI}
\end{figure}

\begin{figure}
  \includegraphics[width=\linewidth]{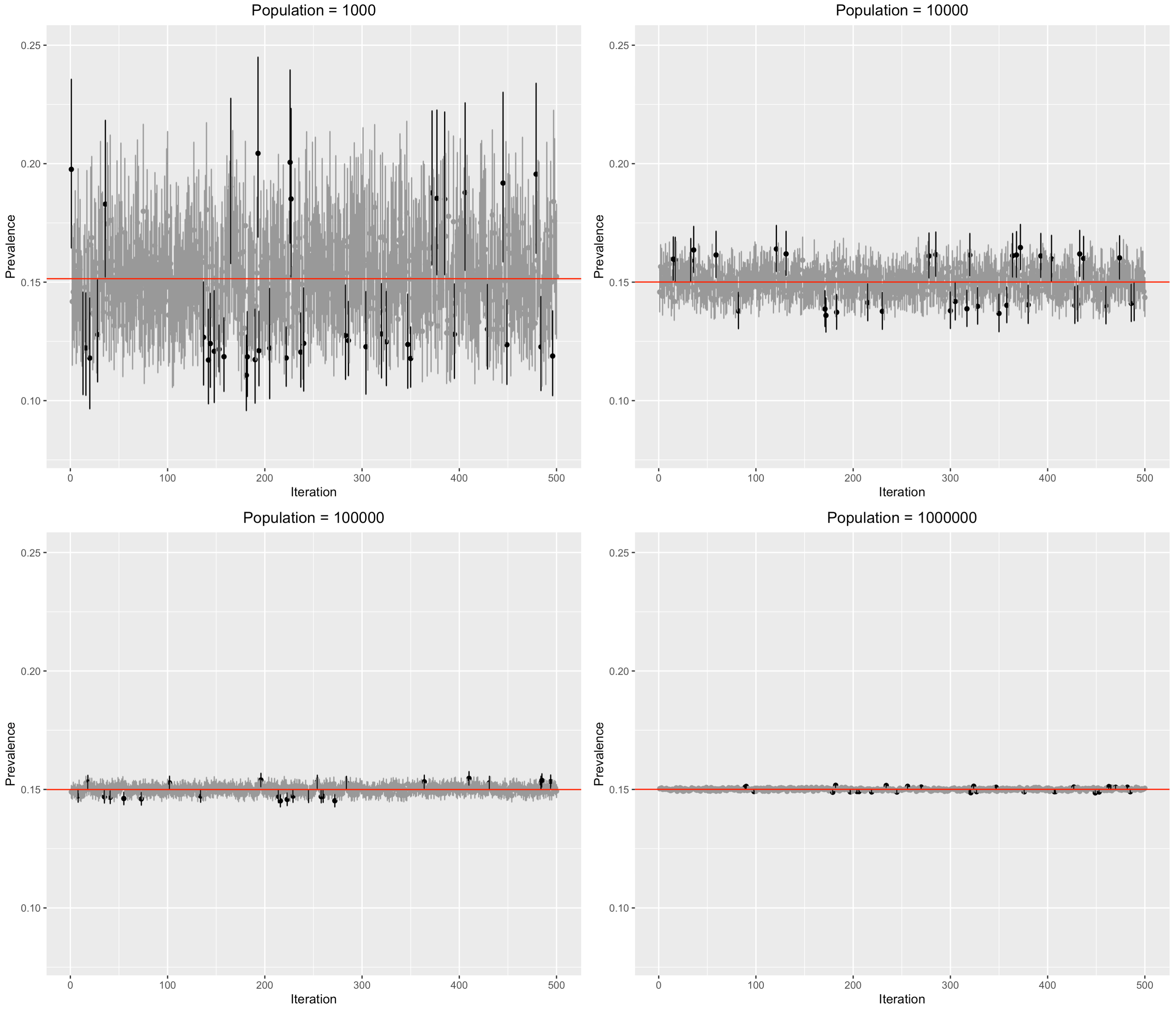}
  \caption{Confidence interval plots for 500 simulations and increasing population sizes under the \textsc{mar}~scenario 2, with $\rho_1=0.2$ and $p_0=0.15$.}
  \label{fig:largeCI}
\end{figure}

Remark \ref{RemCI} is used to estimate $\sigma_{p_0}^2=V_3/N$. Figures~\ref{fig:smallCI} and \ref{fig:largeCI} show 95\% CIs for $p_{0}$ over 500 realizations of the simulations for increasing $N$ for each scenario with the red dashed lines indicating the true value of $p_{0}$.   Table~\ref{table:coverage} gives the empirical coverage probabilities for these scenarios.  

\begin{table}[h!]
\centering
\caption{Fraction of confidence intervals $\mbox{CI}_{p_0}$ for the two \textsc{mar}~scenarios of Figures \ref{fig:smallCI} and \ref{fig:largeCI} that capture the true prevalence $p_0$ out of 500 runs of the simulation. The nominal coverage is $1-\alpha = 0.95$}
\label{table:coverage}
 \begin{tabular}{||c c c ||} 
 \hline
 Population size $N$ & $\rho_1$=0.1, $p_0$=0.05 &  $\rho_1$=0.2, $p_0$=0.15 \\ [0.5ex] 
 \hline\hline
1000 & 0.85 & 0.916  \\  [1ex] 
10000 & 0.898 & 0.928  \\  [1ex]
100000 & 0.912 & 0.95  \\  [1ex]
1000000 & 0.916 & 0.942  \\  [1ex]
 \hline
 \end{tabular}
\end{table}

\section{Discussion}\label{Sec:Disc}


The results of this paper can be extended in various ways. A first extension is to consider prevalence estimation in different laboratories $l=1,\ldots,L$, so that the population is divided into subpopulations $\mathcal X_{lsi}$ with different combinations of laboratories $l$, symptoms $s$ and infection status $i$. In this context, the prevalence can be made to not only depend on symptoms but also on labs. That is,
$$
	p_{0ls} = \frac{|\mathcal X_{ls0}|}{|\mathcal X_{ls0}|+|\mathcal X_{ls1}|},
$$
 within each lab-symptom stratum $(l,s)$, reflects that different labs have different testing procedures.

A second extension is to add errors in testing as in \citep{ZhouDiazRao2022}. In this scenario the number of {\it observed} individuals in each group $\mathcal X_{si}$ are different than $N_{Tsi}$, those actually {\it sampled} from that group.

A third extension, arising naturally from this article, is to consider \textsc{mnar}~settings more systematically. As mentioned in Section  \ref{Sec:ActInfo}, this is the most challenging situation.

\section*{Appendix: Proof of Theorem \ref{Thrm:As}}\label{Sec:Proofs}

To prove \eqref{hpConv}, write 
\begin{align}\label{hpminp}
	\hat p - p &= \sum_{s=0}^{S-1} \tilde \rho_s (\hat p_{0s}-p_{0s}) + \sum_{s=0}^{S-1} (\rho_{Ts}-\tilde \rho_s)p_{0s} + \sum_{s=0}^{S-1} (\rho_{Ts} - \tilde \rho_s)(\hat p_{0s} - p_{0s}).
\end{align}

Each term in the right-hand side of \eqref{hpminp} is now analyzed. From \eqref{hp0s}-\eqref{NTs1Distr} and properties of the hypergeometric distribution \citep[see, for instance,][]{Gut1995}, 
\begin{equation}\label{hp0sConv}
	N^{1/2}(\hat p_{0s}-p_{0s}) \longrightarrow_\mathcal L N\left(0,\rho_s^{-1} \frac{1-\pi_s}{\pi_s} p_{0s}(1-p_{0s})\right)
\end{equation}
as $N\to\infty$, for $s=0,\ldots,S-1$. And from \eqref{hp0s}, the members of $\{\hat p_{0s}\}_{s=0}^{S-1}$ are asymptotically independent. Together with \eqref{hp0sConv}, this implies 
\begin{equation}\label{V1}
	N^{1/2} \sum_{s=0}^{S-1} \tilde \rho_s (\hat p_{0s}-p_{0s}) \longrightarrow_\mathcal L N\left(0, V_1\right),
\end{equation}
where
$$
	V_1 = \sum_{s=0}^{S-1}\tilde \rho_s^2\rho_s^{-1} \frac{1-\pi_s}{\pi_s} p_{0s}(1-p_{0s}) 
	=  \frac{\sum_{s=0}^{S-1}\rho_s\pi_s(1-\pi_s)p_{0s}(1-p_{0s})}{\left(\sum_{s=0}^{S-1} \rho_s \pi_s \right)^2},
$$
and in the second step \eqref{trhosi} was used. 

As for the second term of \eqref{hpminp}, the number of tested individuals with symptoms $s$ is binomially distributed,
$$
	N_{Ts} \sim \mbox{Bin}(N\rho_s,\pi_s),
$$
for $s=0,\ldots,S-1$. Writing $N_{Ts}/N=\rho_s\pi_s + \varepsilon_s$, \eqref{rhoTs} yields that 
\begin{align*}
	\rho_{Ts} &= \frac{\rho_s\pi_s + \varepsilon_s}{\sum_{r=0}^{S-1}(\rho_r\pi_r + \varepsilon_r)} = \tilde \rho_s 
	+ \frac{\varepsilon_s}{\sum_{r=0}^{S-1} \rho_r\pi_r} - \frac{\rho_s\pi_s\sum_{r=0}^{S-1}\varepsilon_r}{\left(\sum_{r=0}^{S-1} \rho_r\pi_r\right)^2} \nonumber\\
	&+ \frac{\sum_{r=0}^{S-1} \varepsilon_r}{\sum_{r=0}^{S-1} \rho_r \pi_r} \left[ \tilde \rho_s - \frac{\rho_s \pi_s + \varepsilon_s}{\sum_{r=0}^{S-1} (\rho_r \pi_r + \varepsilon_r)}\right],
\end{align*}
and the last term on the right-hand side is $o_p\left(N^{-1/2}\right)$. So the second sum of \eqref{hpminp} reads 
$$
	\sum_{s=0}^{S-1} (\rho_{Ts}-\tilde \rho_s)p_{0s} = \frac{\sum_{s=0}^{S-1}p_{0s}\varepsilon_s}{\sum_{s=0}^{S-1} \rho_s\pi_s}
	 -  \frac{\sum_{s=0}^{S-1}\rho_s\pi_s p_{0s}\sum_{s=0}^{S-1}\varepsilon_s}{\left(\sum_{s=0}^{S-1} \rho_s\pi_s\right)^2} + o_p\left(N^{-1/2}\right).
$$
This gives
\begin{equation}\label{V2}
	N^{1/2}\sum_{s=0}^{S-1} (\rho_{Ts}-\tilde \rho_s)p_{0s} \longrightarrow_\mathcal L N\left(0, V_2\right),
\end{equation}
where
\begin{align*}
	V_2 &= \sum_{s=0}^{S-1} \rho_s\pi_s(1-\pi_s) \left[ \frac{p_{0s}}{\sum_{r=0}^{S-1} \rho_r\pi_r} 
		-  \frac{\sum_{r=0}^{S-1}\rho_r\pi_r p_{0r}}{\left(\sum_{r=0}^{S-1} \rho_r\pi_r\right)^2}\right]^2\\
	&= \frac{ \sum_{s=0}^{S-1}\rho_s\pi_s(1-\pi_s)(p_{0s}-\tilde p_0)^2}{\left(\sum_{s=0}^{S-1} \rho_s\pi_s\right)^2},
\end{align*}
and
$$
	\tilde p_0 = \frac{\sum_{s=0}^{S-1}\rho_s\pi_s p_{0s}}{\sum_{s=0}^{S-1}\rho_s\pi_s}
$$
is a weighted average of $p_{00},\ldots,p_{0,S-1}$. 

From the definitions, the first two terms in the right-hand side of \eqref{hpminp} are asymptotically independent. Moreover, the last term of \eqref{hpminp} is $o_p\left(N^{-1/2}\right)$, since $\hat{p}_{0s}-p_{0s}=O_p\left(N^{-1/2}\right)$ according to \eqref{hp0sConv}, and $\rho_{Ts}-\tilde{\rho}_s=O_p\left(N^{-1/2}\right)$ according to the second displayed equation above \eqref{V2}. Equation \eqref{hpConv} therefore follows from \eqref{hpminp}, \eqref{V1}, and \eqref{V2}, by summing the asymptotic variances of the latter two formulas.

To prove \eqref{hp0Conv},  \eqref{hpi02} and \eqref{bap0N} are first used, so that the estimation error of $\hat p_0$ is expressed as
\begin{equation}\label{R3}
	R_3 = \hat p_0 - \bar p_{0N} = \sum_{s=0}^{S-1} \hat \rho_s (\hat p_{0s}-p_{0s}).
\end{equation}
By \eqref{bap0} and a similar argument to the one that led to \eqref{V1}, 
\begin{equation}\label{R3Conv}
	N^{1/2}R_3 \longrightarrow_\mathcal L N(0,V_3),
\end{equation}
with
$$
V_3 = \sum_{s=0}^{S-1}\bar{\rho}_s^2\rho_s^{-1} \frac{1-\pi_s}{\pi_s} p_{0s}(1-p_{0s}). 
$$

Only \eqref{hI+TConv} remains to be proven. To this end,  
\begin{equation}
\hat I_T^+ = I_T^+ - \log \frac{\bar p_{0N}}{p_0} + \log \frac{\hat p}{p} - \log \frac{\hat p_0}{\bar p_{0N}}. 
\label{hI+T2}
\end{equation}
Consequently, by a Taylor expansion of the logarithmic function around 1, 
\begin{equation}\label{hIT+Exp}
	\hat I_T^+ - \left(I_T^+ - \log \frac{\bar p_{0N}}{p_0}\right) = \frac{R_1 + R_2}{p} - \frac{R_3}{\bar p_{0N}} + o_p\left(N^{-1/2}\right),
\end{equation}
where $R_1 = \sum_{s=0}^{S-1} \tilde \rho_s (\hat p_{0s}-p_{0s})$ and $R_2 = \sum_{s=0}^{S-1} (\rho_{Ts}-\tilde \rho_s)p_{0s}$ are the first two terms on the right hand side of \eqref{hpminp}.

In analogy with \eqref{V1}, \eqref{V2} and \eqref{R3Conv}, $N^{1/2}(R_1,R_2,R_3)$ can be shown to converge weakly: 
\begin{equation}\label{R123}
	N^{1/2}(R_1,R_2,R_3) \longrightarrow_\mathcal L N\left((0,0,0),\left(\begin{array}{ccc} V_1 & 0 & V_4 \\ 0 & V_2 & 0 \\ V_4 & 0 & V_3 \end{array}\right)\right),
\end{equation}
where
$$
V_4 = \sum_{s=0}^{S-1}\tilde \rho_s\bar{\rho}_s\rho_s^{-1} \frac{1-\pi_s}{\pi_s} p_{0s}(1-p_{0s}) 
= \frac{\sum_{s=0}^{S-1}\bar{\rho}_s (1-\pi_s)p_{0s}(1-p_{0s}) }{\sum_{s=0}^{S-1} \rho_s \pi_s }.
$$
The proof of \eqref{hI+TConv} is finalized using \eqref{hIT+Exp} and \eqref{R123} and the fact that $\bar p_{0N}\longrightarrow_p \bar p_0$ as $N\to\infty$, which follows from \eqref{hrhosConv}.

\bibliographystyle{plainnat}

\bibliography{/Users/daniela.diaz/Documents/Research/daangapaBibliography.bib}
\end{document}